\documentclass[twocolumn,tighten]{aastex62}
\usepackage{graphicx}
\usepackage{dcolumn}   
\usepackage{amssymb,amsmath,framed,mathtools}

\usepackage{bm}
\expandafter\ifx\csname package@font\endcsname\relax\else
 \expandafter\expandafter
 \expandafter\usepackage
 \expandafter\expandafter
 \expandafter{\csname package@font\endcsname}
\fi
\def\bq{\begin{equation}}
\def\eq{\end{equation}}
\def\bqy{\begin{eqnarray}}
\def\eqy{\end{eqnarray}}

\begin{document}
\title{\large{Active Galactic Nuclei: Boon or Bane for Biota?}}

\correspondingauthor{Manasvi Lingam}
\email{manasvi.lingam@cfa.harvard.edu}

\author{Manasvi Lingam}
\affiliation{Institute for Theory and Computation, Harvard University, Cambridge MA 02138, USA}

\author{Idan Ginsburg}
\affil{Institute for Theory and Computation, Harvard University, Cambridge MA 02138, USA}

\author{Shmuel Bialy}
\affil{Institute for Theory and Computation, Harvard University, Cambridge MA 02138, USA}

\begin{abstract}
Active Galactic Nuclei (AGNs) emit substantial fluxes of high-energy electromagnetic radiation, and have therefore attracted some recent attention for their negative impact on galactic habitability. In this paper, we propose that AGNs may also engender the following beneficial effects: (i) prebiotic synthesis of biomolecular building blocks mediated by ultraviolet (UV) radiation, and (ii) powering photosynthesis on certain free-floating planets and moons. We also reassess the harmful biological impact of UV radiation originating from AGNs, and find that their significance could have been overestimated. Our calculations suggest that neither the positive nor negative ramifications stemming from a hypothetical AGN in the Milky Way are likely to affect putative biospheres in most of our Galaxy. On the other hand, we find that a sizable fraction of all planetary systems in galaxies with either disproportionately massive black holes ($\sim 10^{9-10}\,M_\odot$) or high stellar densities (e.g., compact dwarf galaxies) might be susceptible to both the beneficial and detrimental consequences of AGNs, with the former potentially encompassing a greater spatial extent than the latter.\\ 
\end{abstract}

\section{Introduction} \label{SecIntro}
Active Galactic Nuclei (AGNs) are among the most luminous objects in the Universe \citep{Rees,Ant93,Krol99}. They originate from the accretion of matter onto supermassive black holes (SMBHs) and can emit electromagnetic radiation at rates close to the maximum equilibrium value, the so-called Eddington luminosity.\footnote{In actuality, our analysis applies to black holes at the centers of galaxies, which are not always ``supermassive'' ($\gtrsim 10^5\,M_\odot$) in nature. However, for the sake of brevity, we employ the notation SMBHs henceforth for all central black holes.} Our theoretical and observational understanding of AGNs has progressed considerably over the past couple of decades \citep{FF05,Ho08,HB14,Net15,KP15,PAA17,HA18}. In contrast, their impact on habitability and putative extraterrestrial biospheres has remained comparatively unexplored, with some exceptions delineated below.

\citet{Cla81} was probably the first to propose that cosmic rays emitted by a hypothetical AGN may explain the apparent absence of complex extraterrestrial life in the Milky Way. \citet{LaV83,LaV87} argued, on a related note, that the Galactic center has been sporadically active on timescales of $\sim 10^4$ years, thus emitting high fluxes of energetic electrons (up to $10^5$ times the background value) leading to abrupt changes in Earth's climate and potential mass extinctions. \citet{Gonz05} reviewed the role of AGNs in regulating habitability, and pointed out the possibility that the X-ray fluxes at the Earth arising from a hypothetical AGN in the Milky Way could be comparable to that contributed by the active Sun; this result is also consistent with the subsequent analysis by \citet{ASC14}.

A number of recent studies have been devoted to examining the effects of electromagnetic radiation emitted by AGNs on habitability.\footnote{While AGNs are indubitably important in this regard, their contribution to the deleterious radiation budget at any cosmic epoch is potentially a couple of orders smaller compared to radiation from supernovae \citep{DWC16}.} A large fraction of the radiation from AGNs is emitted as X-rays and ultraviolet (XUV) photons that can heat planetary atmospheres and enhance the rates of atmospheric escape. A detailed analysis by \citet{FL18} concluded that $\sim 10\%$ of all planets in the Universe could have lost as much atmospheric mass as that of the Earth's atmosphere. If the SMBH at the center of the Milky Way, Sagittarius A* \citep{EHK17}, emitted radiation close to the Eddington limit, \citet{BT17} determined that planets $\lesssim 1$ kpc from the AGN are susceptible to losing atmospheric mass equivalent to Earth's atmosphere (see also \citealt{WKB19}). Furthermore, this study also proposed that putative extraterrestrial organisms might suffer damage due to XUV radiation at sub-kpc scales. 

On the other hand, at distances $\lesssim 20$ pc, \citet{CFL18} found that AGNs could have a beneficial effect by stripping sub-Neptunes of thick hydrogen-helium atmospheres that are uninhabitable, thereby converting them into rocky super-Earths. Note, however, at these distances, the X-ray fluxes emitted by AGNs are appreciable \citep{Gonz05}, owing to which many recent studies of galactic or cosmic habitability have posited that only planets located $\gtrsim 10$-$100$ pc from active SMBHs are conducive to sustaining life \citep{DWC16,GH16,SHL18}.

Hitherto, most of the analyses have been dedicated to exploring the negative consequences of AGNs for habitability. Our purpose in this paper is twofold. First, we will explicate some of the positive consequences associated with AGNs (during the quasar phase) for the synthesis of prebiotic compounds (Section \ref{SecUVOOL}), as well as enabling photosynthesis on free-floating planets up to a certain distance from the SMBH (Section \ref{SecPho}). Second, we will revisit the negative effects of AGNs on biota in Section \ref{UVCMLife} by accounting for some crucial factors that were not considered explicitly in prior work. In both cases, we will suppose that the planets under consideration have sufficiently thick atmospheres to avoid being completed eroded due to AGNs.\footnote{Needless to say, in the absence of an atmosphere, much of our subsequent discussion will be rendered inapplicable.}

\section{Ultraviolet fluxes for prebiotic chemistry}\label{SecUVOOL}
We examine the critical distance over which AGNs can stimulate prebiotic chemistry as well as the timescales over which these reactions could occur. Henceforth, we operate under the premise that the UV radiation from the AGN is not strongly attenuated by the dusty torus surrounding the AGN \citep{UP95,HA18}, as well as dust and gas in the cores of galaxies. In analyzing the impact of AGNs on habitability, \citet{BT17} argued that this assumption may not be particularly problematic when dealing with X-rays and UV radiation because the ensuing results must be lowered by an extra factor of $\exp\left(-\tau\right)$, where $\tau$ is the optical depth that obeys $\tau \lesssim 1$. Hence, our results might be altered only by a factor of order unity in this scenario, owing to which we shall proceed without explicitly incorporating this correction.

\subsection{Critical galactic distance for prebiotic chemistry}\label{SSecCDPC}
The bolometric luminosity ($L$) of AGNs is modeled as
\begin{equation}\label{BolLum}
  L \approx 1.3 \times 10^{38}\,\mathrm{erg\,s^{-1}}\,\varepsilon_\mathrm{Edd} \left(\frac{M_{BH}}{M_\odot}\right),
\end{equation}
where $M_{BH}$ denotes the mass of the SMBH and $\varepsilon_\mathrm{Edd}$ is the Eddington ratio. We leave $\varepsilon_\mathrm{Edd}$ unspecified for the time being, but its value is typically close to unity during the quasar phase \citep{MRG04,ACG18}. It must be noted, however, that Eddington ratios of $\varepsilon_\mathrm{Edd} \sim 0.01$-$0.1$ have been documented for several AGNs \citep{LCV10,SE10,HB14,MCM18} and $\varepsilon_\mathrm{Edd} \gtrsim 10$ is also possible \citep{SIH16,CWL17,BV17,WAB18}. 

However, we are not interested in the bolometric luminosity, but in the fraction $\eta_{OL}$ (OL denotes origin of life) of radiation that is emitted in the wavelength range $200 < \lambda < 280$ nm. This is because laboratory experiments pertaining to ``cyanosulfidic metabolism'' \citep{Suth16,Suth17} - a reaction network that yields the precursors of amino acids, nucleic acids, and lipids from a common set of reactants under potentially plausible geochemical conditions - have been conducted in this regime \citep{PCR15,XRT18}. Therefore, the energy flux $\Psi_{OL}$ in the above wavelength range received by an object is
\begin{equation}\label{FluxOOL}
    \Psi_{OL} \approx \frac{\eta_{OL} L}{4\pi d^2},
\end{equation}
with $d$ representing the distance of the object from the AGN. In deriving this formula, we have modeled the emission as being isotropic, but this represents an idealization because of the dusty torus around the AGN \citep{UP95}, as noted earlier, that can break the assumption of spherical symmetry. We calculate $\eta_{OL}$ by making use of the spectral energy distribution presented in \citet[][Figure 1]{VF09}. We find that $\eta_{OL} \approx 2.1 \times 10^{-2}$. Note that the spectral energy distribution behaves roughly as a power-law with exponent $-1$ at lower photon energies \citep{VF09,BT17}.

Next, we make use of the fact that a minimum energy flux is necessary for UV-mediated prebiotic reactions involving sulfur dioxide as the stoichiometric reductant to dominate over ``dark'' reactions that take place in the absence of UV light \citep{RXT18}. Following \citet[][Equation 33]{RXT18}, we find that the photon number flux must exceed the threshold of $5.44 \times 10^{12}$ photons cm$^{-2}$ s$^{-1}$. In order to convert this into the energy flux, we note that the energy of an individual photon only varies by a factor of $1.4$ across this wavelength range. Therefore we treat the photon energy as being approximately constant over this range by making use of the arithmetic mean of the two energies, namely, $\bar{E}_{UV} = 8.5 \times 10^{-12}$ erg. Hence, the critical energy flux translates to $\Psi_c \approx 46.3$ erg cm$^{-2}$ s$^{-1}$. The corresponding distance ($d_O$) at which $\Psi_c$ is attained is calculated from (\ref{FluxOOL}). After simplification, we end up with
\begin{equation}\label{dcOOL}
    d_O \approx 2.2 \times 10^{-2}\,\mathrm{pc}\,\,\varepsilon_\mathrm{Edd}^{1/2}\,\sqrt{\frac{M_{BH}}{M_\odot}}.
\end{equation}
We have plotted $d_O$ as a function of the mass of the SMBH in Figure \ref{FigA}.

If we choose $\varepsilon_\mathrm{Edd} \approx 1$ and $M_{BH} \approx 4 \times 10^6\,M_\odot$ based on the mass of Sagittarius A* \citep{GEG10,BGS16}, we obtain $d_O \approx 44$ pc. Such a value may appear too small for contributing significantly to prebiotic synthesis in our Galaxy. However, note that the stellar density in central regions is very high. Using the Nuker model for the stellar density profile near the Galactic center \citep{SGD18}, we find that as many as $\sim 10^8$ stars might be enclosed within this region.\footnote{The total number of stars in the Milky Way is on the order of $10^{11}$; thus, the number in the central region is comparatively small, but not altogether negligible.} Moreover, the average distances between stars are usually on the order of $10^4$ to $10^5$ AU (i.e., comparable to the Oort cloud) close to the Galactic center, thereby making it much more easier for the exchange of prebiotic compounds and possibly microorganisms as well \citep{CFL18,GLL18}.

Perhaps more importantly, the masses of SMBHs vary widely. The masses of some SMBHs in the Universe, such as the one residing in NGC 4889 \citep{MM12}, are $\sim 10^4$ times higher than Sagittarius A*. The compact lenticular galaxy NGC 1277 might constitute another example \citep{VGG12}; see, however, \citet{GDS16}. For such galaxies, the value of $d_O$ would be raised by roughly two orders of magnitude with respect to the Milky Way. In general, the mass of the SMBH and the stellar mass of the host spheroid (bulge) or the galaxy are proportional to each other \citep{HR04,MM13}, but this plot is characterized by considerable scatter \citep{FCD06,Fab12,BCC12,Gra16}. In this context, note that the $M$-sigma relationship used to infer $M_{BH}$ may break down for low-mass galaxies, as they are anticipated to host under-massive central black holes \citep{PLMM18,NSN19}. Hence, apart from certain exceptions discussed below, the spatial extent of the zone for prebiotic chemistry will be suppressed for low-mass galaxies due to the much lower values of $M_{BH}$.\footnote{This conclusion is also broadly applicable to the zones for photosynthesis (Section \ref{SecPho}) and biological extinctions driven by UV radiation (Section \ref{UVCMLife}).} For example, if we consider the low-mass galaxy NGC 205 with $M_{BH} \approx 6.8 \times 10^3\,M_\odot$ \citep{NSN19}, we end up with $d_O \approx 1.8\,$ pc for the canonical choice of $\varepsilon_\mathrm{Edd} = 1$ after using (\ref{dcOOL}).

Most of the observed galaxies with anomalously high SMBH masses may have ended up as outliers due to a combination of tidal stripping and early formation times \citep{BSB16,VDPD,VBB19}. When it comes to certain compact dwarf galaxies, it is plausible that a sizable fraction (perhaps approaching $\sim 0.1$) of their stars and planetary systems could lie within $d \leq d_O$. The ultra-compact dwarf galaxy (UCD) M60-UCD1 appears to constitute one such specific example. Using $M_{BH} \approx 2.1 \times 10^7\,M_\odot$ for this galaxy \citep{SVM14}, we obtain $d_O \approx 100$ pc. In contrast, its effective radius is only $24$ pc \citep{SVM14}, thereby indicating that a substantial fraction of its planetary systems will ostensibly lie within the region $d \leq d_O$. While M60-UCD1 represents an extreme case of a UCD, it nevertheless underscores our earlier point because these galaxies typically possess effective radii on the order of $10$ pc \citep{BRSF}.\footnote{Recall, however, that UCDs are suspected to represent the tidally stripped remnants of galaxies with masses on the order of $10^9\,M_\odot$ \citep{ASDB}; it is not entirely clear whether UCDs can host AGNs after having evolved to this state.}

\subsection{Timescales for prebiotic chemistry}
If we accept the premise that AGNs can stimulate prebiotic chemistry, we are confronted with an important question: how long are they active? Theoretical models indicate that the characteristic lifetime might be well-approximated by the Salpeter time, i.e. the timescale over which the mass of the SMBH is approximately doubled under the assumption of Eddington-limited accretion. The Salpeter time ($t_S$) is independent of the black hole's mass \citep{Shen13}, and is expressible as
\begin{equation}
   t_S \approx 4.5 \times 10^7\,\mathrm{yrs}\,\,\left(\frac{1}{\varepsilon_\mathrm{Edd} \left(1-\epsilon_{BH}\right)}\right)\left(\frac{\epsilon_{BH}}{0.1}\right), 
\end{equation}
where $\epsilon_{BH}$ is the radiative efficiency that is often approximately equal to $0.1$ \citep{Cop03}. If we specify $\varepsilon_\mathrm{Edd} \sim 0.5$, we find that the Salpeter time is on the order of $10^8$ yrs; note that accretion timescales as high as $\sim 10^9$ yrs have been obtained for some models \citep{TVN17}. This result is consistent with more sophisticated analyses that have yielded an average AGN lifetime of $\sim 4.5 \times 10^8$ yrs provided that $M_{BH} < 10^8\,M_\odot$ and $\sim 1.5 \times 10^8$ yrs for $M_{BH} > 10^8\,M_\odot$ \citep{MRG04}; in some instances, a total lifetime of $\sim 10^9$ yrs for the active phase is feasible.

The earliest unambiguous evidence for life on Earth is attributable to carbon isotopic ($\delta^{13}$C) signatures and stromatolite-like structures from the Isua supracrustal belt in west Greenland that date to $3.7$ Ga \citep{PT18}, but other ambiguous signatures have been dated to $4.1$-$4.3$ Ga \citep{BB15,DPG17}.\footnote{Note that $1$ Ga is the conventional nomenclature for one billion yrs ago (i.e., $1$ Gyr ago).} Thus, based on empirical evidence, we can say with some degree of confidence that life originated within $8 \times 10^8$ yrs after the Earth became habitable with the possible upper bound being as small as $\sim 2 \times 10^8$ yrs. Based on geological considerations, some authors have proposed that the origin of life may have required only $10^7$ to $10^8$ yrs \citep{OF89,LM94,LiMa17} although these arguments have been critiqued by others \citep{Org98}. 

An abiogenesis timescale on the order of $10^8$ yrs is also seemingly consistent with recent phylogenetic analyses that favor life's emergence shortly after the Moon-forming impact at $\sim 4.5$ Ga \citep{BPC18}; on the other hand, molecular clocks are subject to increasing errors as one moves toward earlier epochs. Yet, even if life actually did take $\lesssim 10^8$ yrs to originate on Earth, there is no guarantee \emph{a priori} that it would take the same time elsewhere \citep{ST12}. Nonetheless, based on the evidence we have to date, the timescale of AGN activity might suffice to initiate UV-driven prebiotic chemistry and pave the way for abiogenesis on some planets resembling Hadean Earth.

\section{Powering photosynthesis}\label{SecPho}
As light represents an abundant source of energy, it is not surprising that organisms evolved photosynthesis early in the evolutionary history of Earth \citep{Knoll15,KN17}. The majority of carbon fixation, i.e., the biological synthesis of organic compounds from inorganic carbon sources, on Earth presently takes place via oxygenic photosynthesis that is accompanied by the release of O$_2$ as a product \citep{FBR98}. Given that AGNs emit copious fluxes of electromagnetic radiation, it is natural to ask whether they can power photosynthesis (especially oxygenic photosynthesis). Our analysis is particularly relevant for free-floating planets (or moons) since they do not receive radiation from the host star, but may otherwise possess temperatures and conditions amenable to habitability \citep{Ste99,LA00,DS07,Bad11,Manas19}. Based on quasar microlensing, it was recently determined that the number of free-floating objects with sizes ranging from the Moon to Jupiter is $\sim 2\times 10^3$ per main-sequence star \citep{DG18}. If we posit this ratio also applies in the vicinity of the Galactic center, we find that $\sim 10^{11}$ free-floating objects of this size range might exist within the volume encompassed by $d \sim 100$ pc (see Section \ref{SecUVOOL}).

We will therefore study how AGNs could enable the evolution and sustenance of photoautotrophs (presupposing abiogenesis had already been initiated through some pathway) on free-floating worlds as well as the timescales associated with its emergence.

\subsection{Critical galactic distance for photosynthesis}
In order to estimate the range up to which AGNs can facilitate photosynthesis, we note that the photon flux $\Psi_{PAR}$ at distance $d$ is
\begin{equation}\label{FluxPAR}
    \Psi_{PAR} \approx \frac{\eta_{PAR} L}{4\pi d^2},
\end{equation}
where $\eta_{PAR}$ refers to the fraction of electromagnetic radiation emitted by the AGN that is suitable for photosynthesis; such radiation is known as photosynthetically active radiation (PAR). However, in order to compute $\eta_{PAR}$, we are confronted with a couple of stumbling blocks. Even though the range of oxygenic photosynthesis is fairly well established on our planet, it remains poorly constrained on other worlds \citep{WoRa02,KST07}. We will adopt the slightly conservative choice of $\sim 400$-$1000$ nm henceforth for PAR, where the lower bound is set by the inhibition of photosynthesis by UV radiation and the upper bound by the energy required to split a water molecule and release O$_2$ \citep{GW17}. The upper bound is also consistent with certain microbes on Earth that rely upon anoxygenic photosynthesis \citep{KSG07}; in anoxygenic photosynthesis, compounds other than H$_2$O serve as the electron donor (e.g., molecular hydrogen). Using the data from \citet{VF09}, the fraction of total electromagnetic radiation emitted in the above wavelength range by the AGN is chosen to be $\eta_{PAR} \approx 5.7 \times 10^{-2}$. Note that several AGNs are characterized by higher values of $\eta_{PAR}$ that are on the order of $0.1$ \citep{EWM94,HEC14}.

We observe that the photon energy only varies by a factor of $2.5$ over the range specified above, implying that it can be treated as roughly constant. Hence, we will introduce the mean photon energy of $\bar{E}_{PAR} \approx 3.5 \times 10^{-12}$ erg for PAR. Now, in order to determine the minimum energy flux necessary for photosynthesis, we must know the minimum photon flux. Although the latter is also unknown for putative extraterrestrial organisms, in principle a comparatively robust estimate is probably set by biophysical considerations: at least one photon must be incident on the area of the photosynthetic apparatus (e.g., photosystem I or II) over a single temporal cycle involving electron transfer. The minimum empirical bound of $\sim 6 \times 10^{11}$ photons cm$^{-2}$ s$^{-1}$ has been explained through charge recombination in photosystem II, protein turnover and H$^+$ leakage \citep{RKB00}. Therefore, combining these results, the critical PAR energy flux becomes $\Psi_0 \sim 2.1$ erg cm$^{-2}$ s$^{-1}$. Substituting this value into (\ref{FluxPAR}), and solving for the cutoff distance ($d_P$) yields the extent of the ``photosynthesis zone''. By doing so, $d_P$ is given by
\begin{equation}\label{dcPAR}
    d_P \approx 0.17\,\mathrm{pc}\,\,\varepsilon_\mathrm{Edd}^{1/2}\,\sqrt{\frac{M_{BH}}{M_\odot}}.
\end{equation}
A couple of important points are worth highlighting regarding the above formula: (i) the net primary productivity of biospheres at these low light levels is potentially several orders of magnitude smaller with respect to modern Earth, and (ii) photoautotrophs require not only sufficient fluxes of photons at appropriate wavelengths, but also reactants (e.g., CO$_2$), nutrients (e.g., PO$_4^{3-}$) and electron donors \citep{LiMa18,LL19}.

The photosynthesis zone has been plotted as a function of $M_{BH}$ in Figure \ref{FigA}. Substituting the parameters for the Milky Way, we obtain $d_P \approx 0.34$ kpc. This distance is fairly ``large'' since it approximately equals the scale height of the thin disk of our Galaxy. If we consider black holes with masses on the order of $10^{9-10}\,M_\odot$, we find $d_P \gtrsim 10$ kpc. In such instances, large swathes of galaxies would fall within the domain of the photosynthesis zone. As noted in Sections \ref{SecUVOOL} and \ref{UVCMLife}, UCDs are characterized by very high stellar densities, indicating that the photosynthetic zone calculated above could also encompass most stars in these galaxies. 

\subsection{Timescales for the evolution of photosynthesis}
Finally, we must also examine the issue of timescales. Analyses of isotopic ratios on Earth favor the existence of anoxygenic photosynthesis by $\sim 3.5$ Ga \citep{TL04,Ols06,But15}, or possibly even earlier; recall that we posited life had originated by $\gtrsim 3.7$ Ga. The origin of oxygenic photosynthesis remains less tightly constrained, with some proxies (e.g., U-Th-Pb isotopic ratios) placing it as far back as $\gtrsim 3.7$ Ga \citep{RF04}, although unambiguous evidence from microfossils appears to date back only to $1.9$ Ga \citep{LRP14,FHJ16}. Despite this uncertainty, there are multiple grounds for contending that oxygenic photosynthesis may have evolved by $\gtrsim 2.7$ Ga, based on isotopic ratios, phylogenetics and microfossils \citep{Bui08,SSW16}, thus amounting to its emergence by at least several $100$ Myr prior to the Great Oxidation Event at $\sim 2.4$ Ga; see also \citet{KBS16,MaLi18,CSRL}. \emph{In toto}, it is not altogether unreasonable to surmise that both anoxygenic and oxygenic photosynthesis might have evolved over a time span ranging from $10^8$ to $10^9$ yrs after the origin of life on Earth.

Therefore, if one supposes that the rates of molecular evolution and speciation are similar on other worlds, the active phases of some SMBHs may function long enough \citep{MW01,MRG04,TVN17} to permit the evolution of anoxygenic photosynthesis at the minimum. Once the active phase ends, it is plausible that the putative photoautotrophs and their accompanying biospheres could end up becoming extinct. However, the complete loss of biodiversity is not necessarily inevitable as it ultimately depends on the adaptability of the organisms in question.

\section{Harmful effects of ultraviolet radiation on biospheres}\label{UVCMLife}
We explore the critical distance at which UV radiation may cause damage to Earth-like ecosystems, and also briefly discuss the effects of X-rays and gamma rays.

\subsection{Critical galactic distance for UV damage}
Along the lines of Section \ref{SecUVOOL}, we will now compute the fluxes in the UV-A ($315$-$400$ nm), UV-B ($280$-$315$ nm) and UV-C ($122$-$280$ nm) regimes. In order to do so, we must determine the equivalents of $\eta_{OL}$ that are denoted by $\eta_{A}$, $\eta_{B}$ and $\eta_{C}$. Henceforth, the subscripts $A$, $B$ and $C$ correspond to the UV-A, UV-B and UV-C ranges, respectively. The corresponding fluxes are given by
\begin{equation}\label{PhiDef}
    \Psi_j \approx \frac{\eta_{j} L}{4\pi d^2},
\end{equation}
where $j \in \{\mathrm{A, B, C}\}$. By employing the data from \citet{VF09}, we work with $\eta_A \approx 1.5 \times 10^{-2}$, $\eta_B \approx 7.3 \times 10^{-3}$ and $\eta_C \approx 5.2 \times 10^{-2}$. 
The important point to appreciate with respect to the above formula is that it corresponds to the top-of-atmosphere (TOA) fluxes. In general, the shorter the photon wavelength, the greater the destructive effects on biota owing to the higher energy per photon \citep{Dart11}. The deleterious effects associated with UV radiation include the inhibition of photosynthesis and the damage of DNA and other biomolecules.

To leading order, we may exclude the effects of UV-A radiation because its propensity toward inducing carcinogenesis and degradation of cellular activity is much lower compared to UV-B radiation \citep{DG00,MGAB}. Furthermore, $\beta$-carotene in land plants and algae constitutes an effective shield against UV-A radiation \citep{WJ02}. Likewise, bacteria are adept at mitigating the damage by UV-A radiation through the formation of biofilms \citep{EM99}; see also \citet{FPW17}. This leaves us with UV-B and UV-C radiation, thereby necessitating the consideration of two different scenarios. Before tackling them, it is helpful to discuss the impact of a hypothetical AGN situated in the Milky Way on the Earth's biosphere insofar as UV fluxes are concerned. 

By plugging in the parameters for Earth into (\ref{PhiDef}), namely $d \sim 8$ kpc and $M_{BH} \approx 4 \times 10^6\,M_\odot$ \citep{BGS16}, we find that the TOA UV fluxes are $\lesssim 10^{-3}$ erg cm$^{-2}$ s$^{-1}$. In contrast, the corresponding TOA UV fluxes received from modern-day Sun are on the order of $10^3$ to $10^4$ erg cm$^{-2}$ s$^{-1}$. The same conclusion also applies if we compare the results against the TOA UV fluxes from the Sun at different epochs in Earth's history ranging from $3.9$ Ga to the present day \citep{RSK15}. Hence, insofar as the Earth is concerned, the UV fluxes from a hypothetical AGN in the Milky Way are not expected to pose major issues to surface biota since they are several orders of magnitude smaller than the solar UV fluxes. We will now turn our attention to other worlds and consider two cases.

First, we may consider a planet that has an atmosphere akin to Archean Earth at $\sim 3.9$ Ga. Owing to the absence of an ozone layer, nearly all of the TOA UV-B and UV-C fluxes are transmitted to the surface. In this case, the impact of UV-C radiation will be commensurately higher relative to UV-B radiation. The surface UV-C flux incident on the Earth's surface during this epoch was estimated to be approximately $871$ erg cm$^{-2}$ s$^{-1}$ \citep{RSK15}. The fluence of UV-C radiation that kills $90\%$ of \emph{Deinococcus radiodurans}, one of the most radiation-resistant extremophiles, is around $5.5 \times 10^5$ erg cm$^{-2}$ \citep{GOP95}. Thus, in principle, the majority of \emph{D. radiodurans} should become extinct in $\sim 10$ minutes. However, despite these very high UV fluxes, there is clear-cut evidence of life on Earth during the Archean as outlined in Section \ref{SecUVOOL}.

In view of this fact, it is evident that life can exist in high-UV environments. A number of shielding strategies are feasible ranging from photochemical hazes in the atmosphere to subsurface environments underneath soil or several meters below the surface of oceans \citep{LL18}. In addition, complex evolutionary adaptations (e.g., microbial UV sunscreens such as scytonemin, mycosporines and melanins) also render additional radiation resistance \citep{GS13,JLB17}. Therefore, we are led towards an important point: the sensitivity of biota to UV fluxes from AGN depends on the physiology of the organisms themselves. In particular, it is not unreasonable to surmise that any microbes on planets near the Galactic center would have already evolved to function in a high-UV environment arising from the close proximity of supernovae, implying that the additional inclusion of AGN-derived UV radiation is not guaranteed to affect them adversely.

Next, we consider worlds with biospheres and atmospheres resembling that of modern Earth. Due to the presence of an ozone layer, life on Earth's surface experiences a comparatively benign UV environment. For such worlds, the amount of UV-C radiation reaching the planetary surface is nearly zero \citep{RSK15}. This leaves UV-B radiation as the primary threat. It was argued by \citet{MeTh11} that doubling the background (i.e., stellar) UV-B flux at the surface would trigger an extinction-level event because complex biota adapted to live in otherwise temperate environments are not equipped to handle an abrupt and substantial increase in UV-B radiation. 

\begin{figure*}
\begin{centering}
\includegraphics[scale=0.77]{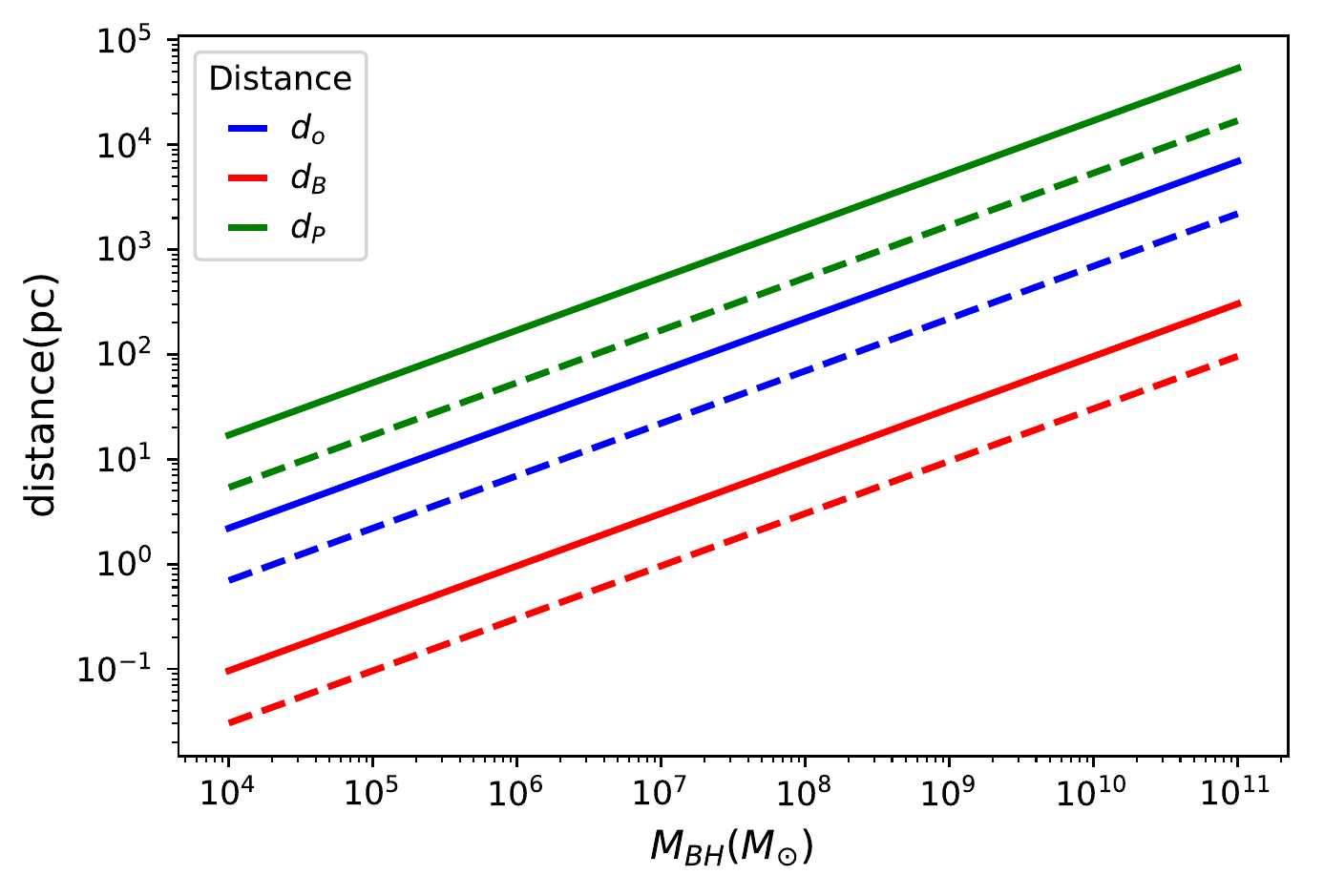}\\
\includegraphics[scale=0.77]{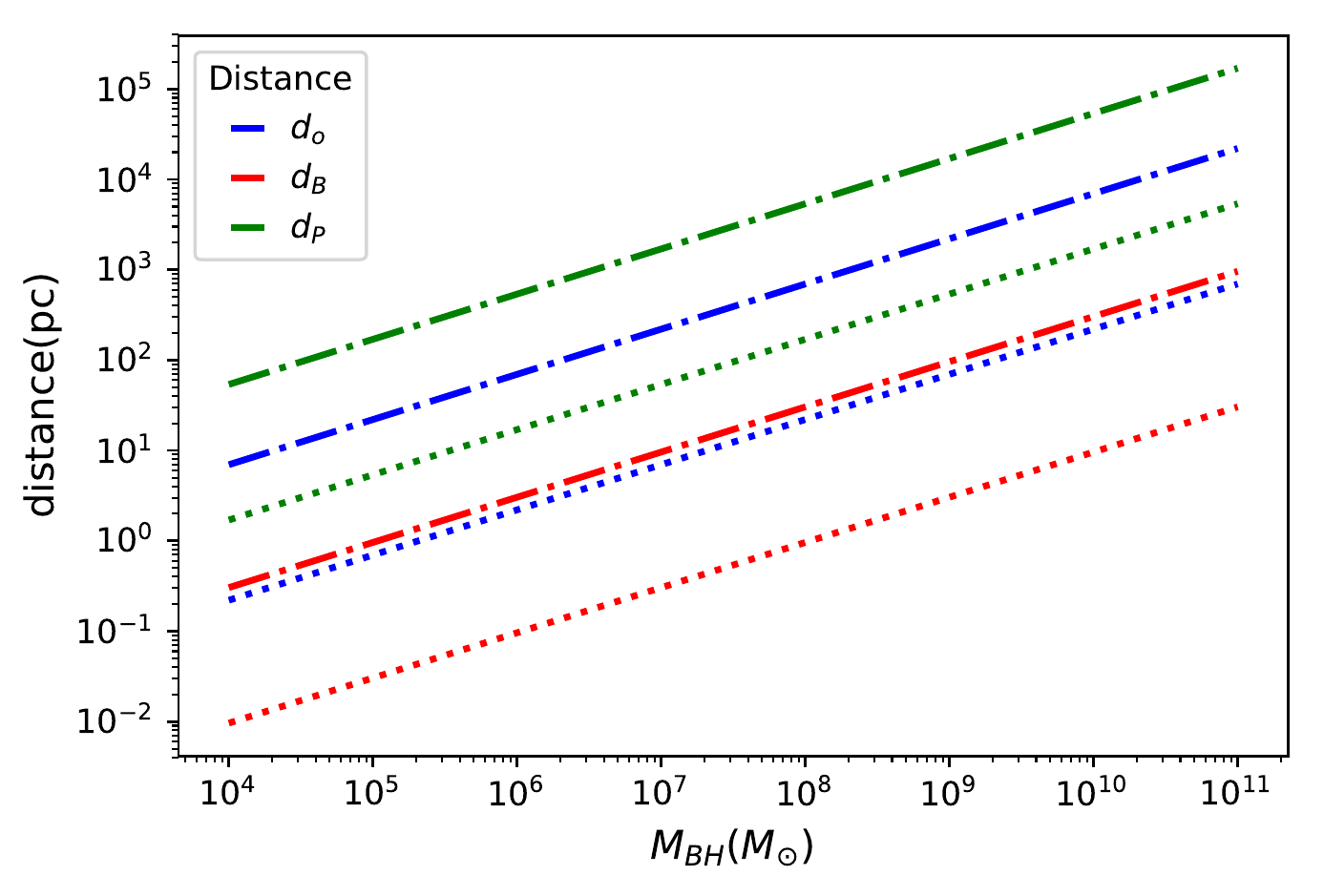}
\caption{The distances at which AGNs enable prebiotic chemistry ($d_O$), facilitates potential extinction events ($d_B$) and permits photosynthesis ($d_P$) are depicted as a function of the mass of the SMBH ($M_{BH}$). The x-axis is truncated at $10^4\,M_\odot$ because this value is quite close to some of the smaller central BHs known to date \citep{MCM18,NSN19,GSD19}, and at $10^{11}\,M_\odot$ as SMBHs are not expected to exceed this theoretical limit at the current age of the Universe \citep{King16,IK16,PNF17}. For Sagittarius A*, we obtain $d_O \approx 44$ pc, $d_B \approx 2$ pc and $d_P \approx 340$ pc after selecting an Eddington factor close to unity. Top panel: the solid and dashed curves correspond to Eddington factors of unity and $0.1$, respectively; note that these two values are consistent with observational constraints that mostly favor average Eddington factors of $\sim 0.1$-$1$ \citep{MRG04}. Bottom panel: the dotted-dashed and dotted curves correspond to Eddington factors of $10$ and $0.01$, respectively; these values are also consistent with observations and theory (see Section \ref{SSecCDPC}).\\}
\label{FigA}
\end{centering}
\end{figure*}

Thus, a sufficient criterion for severe ecological damage can be derived when the TOA UV-B flux contributed by the AGN equals the TOA UV-B flux arising from the host star, as this would suggest that the corresponding surface UV fluxes are comparable to each other as well. The ``typical'' stellar UV-B flux at the TOA is not easy to compute, since it depends on the planet's location as well as the spectral type of the star \citep{LiLo18}. However, in the case of worlds that are similar to modern Earth, we choose a TOA UV-B flux of $8.6 \times 10^3$ erg cm$^{-2}$ s$^{-1}$ \citep{RSK15}. By using (\ref{PhiDef}) along with our estimate for $\eta_B$, we obtain the distance ($d_B$) at which this threshold is satisfied:
\begin{equation}\label{dUVB}
    d_B \approx 9.6 \times 10^{-4}\,\mathrm{pc}\,\,\varepsilon_\mathrm{Edd}^{1/2}\,\sqrt{\frac{M_{BH}}{M_\odot}}.
\end{equation}
In Figure \ref{FigA}, the above distance has been plotted as a function of the SMBH mass.

Thus, applying this formula to the Milky Way, we obtain $d_B \approx 2$ pc. As there are only $\sim 10^6$ stars in this volume, we see that this number is much lower than the critical distance $d_O$ derived for prebiotic chemistry in Section \ref{SecUVOOL}. If we specialize to Earth-like planets in the habitable zone of late M-dwarfs such as Proxima Centauri and TRAPPIST-1, we note that the UV-B flux threshold must be roughly lowered by a factor of $\sim 4 \times 10^{-3}$ relative to Earth \citep{RSK15,RWS17}, implying that $d_B$ will be increased to 
\begin{equation}\label{dUVBMD}
    d_B \approx 1.5 \times 10^{-2}\,\mathrm{pc}\,\,\varepsilon_\mathrm{Edd}^{1/2}\,\sqrt{\frac{M_{BH}}{M_\odot}}.
\end{equation}
For the Milky Way, making use of (\ref{dUVBMD}) leads us to $d_B \approx 30$ pc. Regardless of the spectral classification of the host star, we see that this distance is much smaller than the scale length and height of the Milky Way. As noted earlier, even at $d \sim d_B$, UV radiation from AGNs will not necessarily cause extinction-level events since any putative biota could already be accustomed to the high-radiation environment near the SMBH.

It must, however, be noted that $d_B$ can become fairly large for galaxies with SMBHs on the order of $10^{10}\,M_\odot$. For such galaxies, depending on the TOA UV-B threshold specified, we find that $d_B$ will be raised by as much as two orders of magnitude relative to the Milky Way, i.e., we have $d_B \lesssim 0.2$-$3$ kpc. For an Eddington ratio of unity, it is unlikely that this value of $d_B$ can be exceeded, since the maximal mass of SMBHs is on the order of $10^{10}\,M_\odot$ \citep{NT09,King16,IK16,PNF17}. Next, if we concern ourselves with UCDs, a non-negligible fraction of stars are anticipated to exist within $10$ pc, implying that these galaxies might experience widespread extinction-level events due to high UV doses from AGNs.

\subsection{X-rays and gamma rays}
We begin by observing that photons with wavelengths $\lambda < 170$-$200$ nm are shielded by atmospheric water vapor and carbon dioxide, respectively \citep{RaSa16}. Therefore, when it comes to gamma-ray bursts (GRBs) that emit high fluxes of X-rays and gamma rays, the majority of photons are absorbed by planets with Earth-like atmospheres \citep{MeTh11}. Consequently, we have not modeled the effects of XUV radiation since the majority of these photons are not anticipated to directly reach the surface. Instead, their true impact lies in promoting: (i) atmospheric escape \citep{BT17,FL18,WKB19}, (ii) photolysis of atmospheric H$_2$O leading to the depletion of oceans \citep{FL18,WKB19}, (iii) driving changes in atmospheric chemistry \citep{Dart11}. With regards to the latter, ozone depletion and its accompanying ecological ramifications are potentially prominent \citep{LL17}, if an O$_3$ layer exists in the first place. Detailed atmospheric chemistry models must be employed in the future to properly assess the ensuing biological impact of X-rays and gamma rays emitted by AGNs.

It must be noted that the flux of X-rays received from a hypothetical AGN in the Milky Way at the Earth is on the order of $10^{-3}\,\mathrm{erg\,cm^{-2}\,s^{-1}}$ after using Figure 1 of \citet{VF09} and calculating the flux along the lines of (\ref{PhiDef}); our result is consistent with \citet{ASC14}. In contrast, the equivalent flux from GRBs at kpc distances is on the order of $10^7\,\mathrm{erg\,cm^{-2}\,s^{-1}}$ \citep{Thom09}, but the effects on marine biota are not predicted to be severe \citep{NT16}, due to the much shorter duration of GRBs and the fact that these organisms are protected beneath liquid water. Likewise, based on \emph{GOES} data, the X-ray flux at the Earth from the quiescent Sun is $\sim 10^{-4}\,\mathrm{erg\,cm^{-2}\,s^{-1}}$,\footnote{\url{https://www.swpc.noaa.gov/products/goes-x-ray-flux}} although it increases by a few orders of magnitude during large flares \citep{Benz17}. Hence, with regards to X-rays, the effects on Earth's biota should be minimal, in agreement with \citet{Cla81}.

\section{Conclusion}\label{SecConc}

In this paper, we examined some of the positive and negative consequences of AGNs for life. With regards to the former, we determined that AGNs are capable of powering UV-mediated prebiotic reactions that lead to the synthesis of biomolecular building blocks. We also demonstrated that, because they emit substantial fluxes of visible radiation, it opens the possibility of photosynthesis being powered by AGNs. When it comes to the adverse effects of UV radiation on biota, we found that the significance of AGNs might have been overestimated. In particular, we showed that the ``zone'' where the negative effects become dominant is apparently smaller than the corresponding zones for powering prebiotic chemistry and photosynthesis, as seen from Figure \ref{FigA}; this can also be verified by comparing (\ref{dUVB}) and (\ref{dUVBMD}) with (\ref{dcOOL}) and (\ref{dcPAR}). An implicit caveat regarding our study is worth reiterating: while we have endeavored to keep our treatment as general as possible, some of the biological parameters employed herein were adopted from particular species on Earth, thus implying that the ensuing results are not guaranteed to hold true for other worlds.

Specializing to the case of a hypothetical AGN in the Milky Way, we found that neither positive nor negative phenomena are likely to matter at distances of $\gtrsim 1$ kpc. The zone for UV-induced extinction-level events extends to $\sim 2$-$30$ pc, whereas the equivalent zones for prebiotic chemistry and photosynthesis are $\sim 44$ pc and $\sim 340$ pc, respectively. For galaxies with black holes on the order of $10^4$ to $10^5\,M_\odot$ \citep{CKZ18,MCM18,GSD19}, these zones could be truncated at sub-pc scales; in comparison, note that the Bondi radius for Sagittarius A* is around $0.04$ pc \citep{YN14}. At sub-pc distances, planets may perhaps be subject to complex interactions with the dusty torus as well as strong outflows and winds from the accretion disc \citep{GMP16,GSD18,BMR18}. Hence, close-in planets and moons at $\lesssim 1$ pc might become uninhabitable from continual exposure to hot gas at relativistic speeds. On the other hand, it is important to appreciate that the distance of $1$ pc does not necessarily represent a ``hard'' cutoff, in view of the substantial heterogeneity among active galaxies and the fact that our understanding of their central regions is currently not definitive.

In broad terms, we found that there are two types of galaxies where both positive and negative effects become increasingly important. The first are galaxies with anomalously large SMBHs that are orders of magnitude more massive than Sagittarius A*, potentially reaching values of $10^{9-10}\,M_\odot$. In this case, the aforementioned zones may even reach kiloparsec scales. The second category corresponds to compact dwarf galaxies that are characterized by their high stellar densities, provided that they have active SMBHs. As a result, even if their SMBHs are not very massive, their effective radii are much smaller, in turn implying that a sizable fraction of all their stars ought to fall within the above zones. 

Our analysis of AGNs has focused exclusively on electromagnetic radiation thus far. We anticipate that finding indirect electromagnetic signatures of past hypothetical AGN activity on Earth are almost impossible in all likelihood for two reasons. First, from the standpoint of biological signatures (e.g., elevated extinction rates), we have shown that the deleterious ramifications of a putative AGN on Earth's biosphere were probably negligible, implying that the accompanying signatures in the fossil record would also be minimal. Second, high-energy electromagnetic radiation tends to leave comparatively few direct traces in the geological record \citep{MeTh11}. The same conclusions may also apply to early Mars, but certain aspects of its geological activity, atmospheric composition and climate are not well understood \citep{word16,EAA,RC18,DLML,kite19}.

Future work should endeavor to accurately determine the extent by which Galactic Cosmic Ray (GCR) fluxes are amplified by AGNs, and consequently model the ensuing biological ramifications. Doing so will enable us to achieve progress in uncovering the activity of Sagittarius A* in the past, thereby equipping us with another channel alongside conventional metrics such as X-ray echoes \citep{PMT13} and gamma-ray excesses \emph{\`a la} ``Fermi bubbles'' \citep{PSZ14}. Estimating the increase in GCR fluxes incident on Earth due to a hypothetical AGN in the Milky Way is not a straightforward task because of ambiguities concerning the acceleration mechanisms and spectra of high-energy particles \citep{Krol99} as well as the putative role and significance of planetary magnetic fields \citep{Ling19}.

Consider the fiducial choice wherein the flux of GCRs received by the Earth was doubled due to putative AGN activity. It would probably cause nominal damage to Earth's biosphere because most species, including complex multicellular organisms, are capable of withstanding radiation dose rates a few times higher than the background value \citep{Ken14,TLK16}. Large flares ($\gtrsim 10^{32}$ erg), in particular, transiently boost the number flux of energetic particles penetrating to the surface by many orders of magnitude \citep{Atr17,LDF18}, without necessarily causing large-scale extinctions of biota except perhaps in the most extreme stellar proton events \citep{LL17} accompanying large superflares \citep{NM19}.

A vital point with regards to the preceding issue of GCR fluxes from AGN activity in our Galaxy merits further explication. Measurements of cosmogenic isotopes, especially $^{14}$C and $^{10}$Be, from meteorites and lunar rocks ostensibly enforce fairly stringent constraints over long timescales: the variation in average GCR flux has been within $10\%$ over the past $\sim 10^6$ yrs and less than a factor of $1.5$ during the past $\sim 10^9$ yrs \citep{Gri01,Uso17}. Therefore, either no AGN was functional in the Milky Way during this period or the enhancement in GCR fluxes at the Earth due to such activity was smaller than the above values. In either scenario, the net effect on Earth's biosphere might have been marginal for the reasons elucidated earlier.\\

\acknowledgments
We are grateful to Abraham Loeb, Chelsea MacLeod, and the reviewer for the insightful comments regarding the paper. This work was supported in part by the Breakthrough Prize Foundation, Harvard University, and the Institute for Theory and Computation. 


\end{document}